# Key Body Posture Characteristics of Short-distance Speed Skaters at the Start Based on Artificial Intelligence

Zhang Xuelian[a], Fang Yingjie[b], Liu Hang[c]

a.College of Physical Education, Humanities and Social Sciences, Harbin Sports University.No. 1 Dacheng Street, Nangang District, Harbin.
b.School of Physical Education, Harbin Sport University.No. 1 Dacheng Street, Nangang District, Harbin.
c.Graduate School Harbin Sport University.No. 1 Dacheng Street, Nangang District, Harbin.

**Abstract: Objective** To conduct biomechanical analysis on the starting technique of male short-distance speed skating athletes in China and determine the key factors affecting the effectiveness of the starting movement. **Methods** 13 high-level male short-distance speed skating athletes were selected as the test subjects, and kinematic data were collected using an artificial intelligence video capture and analysis system. The body posture features and their effects on the starting movement performance were analyzed in the three stages of starting preparation, starting, and sprinting. **Results** The post-stability angle, anterior knee angle of the front leg, posterior knee angle of the rear leg, and stride length showed moderate to high positive correlations with the starting speed during the starting preparation stage. The trunk angle showed a high negative correlation with the starting speed. The trunk angle (TO4, TD4, TO6, TD6), hip angle (TO1, TO4, TO6), and knee angle (TD1) showed moderate to high negative correlations with the effectiveness of the starting movement during the starting and sprinting stages. The knee angle (TD2), ice-contact angle (TD2, TD4, TD5, TD6), and propulsion angle (TO1, TO4, TO7) showed moderate positive correlations with the effectiveness of the starting movement. **Conclusion** Stride length, left knee angle, and post-stability angle are the key factors affecting the starting speed. The larger the post-stability angle and left knee angle and the longer the stride length, the faster the starting speed. During the starting and sprinting stages, the smaller the ice-contact angle and propulsion angle, the greater the trunk angle and hip angle changes, the more effective the starting movement.

Speed skating, as one of the most historic competitive sports in the Winter Olympic Games, has consistently held a central position throughout the evolution of the Winter Olympics. Its competitive value carries strategic significance equivalent to that of fundamental events in the Summer Olympics, such as athletics and swimming. The primary objective of this discipline is to complete a predetermined course distance in the shortest possible time. In recent years, China has made remarkable progress in the field of short-track speed skating. However, with the continuous improvement of global speed skating performance, competition among elite athletes has become increasingly intense, with medalists' results differing by mere milliseconds [1]. A systematic review of the literature in speed skating research indicates that athletes' rapid start capability and acceleration performance during the initial phase of the race have a decisive impact on competition outcomes [2,3]. The starting technique, being a crucial winning factor in short-track speed skating [4], directly influences competition results through the precision of its technical details. Therefore, in-depth research into the key techniques of starting in short-track speed skating holds significant theoretical and practical importance.

From a biomechanical perspective, the speed skating start technique refers to the complete technical process in which athletes assume a preparatory position and remain stationary upon receiving the "Set" command from the starter, until the gunshot triggers a rapid start and transition into the skating phase. This technical sequence can be systematically divided into three critical phases: preparatory position, start, and sprint, with the core objective of achieving an optimal transition from a stationary state to the individual's maximum skating speed [4]. The enhancement of starting speed primarily relies on stable support before the start and the coordinated rapid extension of body segments after initiation. Among these, the stability angle serves as a crucial indicator for assessing the stability of the athlete's starting posture, while the trunk angle and ice contact angle effectively reflect upper limb extension capability and forward propulsion. The hip, knee, and ankle joint angles directly characterize lower limb extension capacity, and the push-off angle can be used to evaluate the quality of ice contact support and push-off movements. Scientific start training and competitive practice require meticulous control of each preparatory movement, starting posture, initiation timing, as well as push-off and swing actions [5].

This study, based on an artificial intelligence-powered three-dimensional motion capture system, aims to identify the core factors influencing starting performance in short-track speed skaters through quantitative analysis of key body posture characteristics during the starting process. The findings are expected to provide scientific evidence and theoretical support for optimizing starting techniques.

**1 Research Objectives and Methods**

1.1 Research Objectives

The research subjects consisted of 13 male short track speed skaters . The tests were approved by the Heilongjiang Provincial Ice Training Center and recognized by the relevant coaches. The participants had an average height of $1.79\pm0.06$m, average weight of $74.15\pm5.32$kg, and average record of $37.04\pm0.52$ sec of 500m.The criteria for subject recruitment were: (1) athletes with a national sports master or higher level of athletic ranking; (2) subjects had no injuries to the upper or lower limbs and waist in the past month, with normal physical function; (3) all subjects adopted a standing start position with the left foot in front and the right foot behind.

1.2 Test Process

First, let the athletes understand the purpose and process of the test and sign an informed consent form. After that, the athletes perform a standardized warm-up for about 25 minutes, which includes exercises such as jogging, stretching, and ice skating to make the body slightly sweat. During the formal test, the competition start command process is followed, and video parameters of the starting process are collected, with the experimental data recorded and saved.

1.3 Data Collection

To investigate the characteristics of body posture at the start of short-distance speed skating athletes and their impact on the performance of starting technique, an unmarked point artificial intelligence video acquisition and analysis system (Fastmove Realtime V1.0) was used to test the athletes' starting process. Two high-definition cameras were connected using Vention fiber optic cables, spaced 7.25 meters apart, at a height of 1.2 meters from the ground, 6.5 meters from the marked point of shooting, with an angle of approximately 90 degrees between the main optical axes, a shutter speed of 1/500 seconds, and a sampling frequency of 60Hz, with a resolution of 1920x1080 pixels. The X-axis was determined to be the direction of movement, the Y-axis the left-right direction, and the Z-axis the up-down direction. The test site is shown in Figure 1.

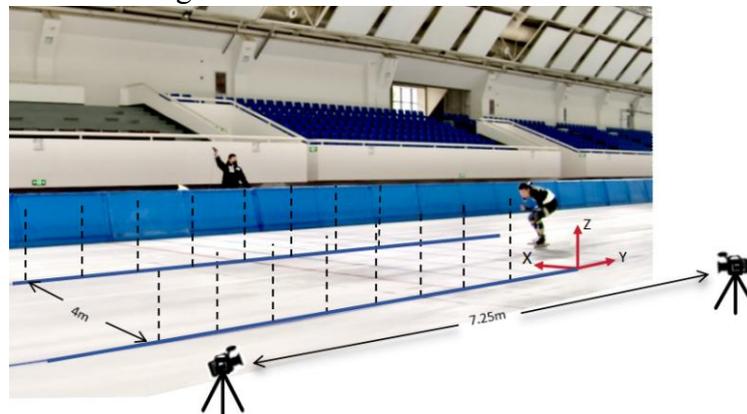

**Fingure 1   The Actual Situation of Test site**

1.4 Data Extraction and Processing

Data processing was performed using the artificial intelligence three-dimensional kinematic analysis system (Fastmove Motion-3D, V1.2.11). First, the system intelligently identified 21 key points on the human body; second, manual corrections were made to "abnormal points"; then, a ButterWorth fourth-order low-pass digital filter was applied for smoothing filtration with a cut-off frequency of 10Hz [6], using AI technology for three-dimensional spatial calibration and ground coordinate transformation, reconstructing the three-dimensional data model; finally, the raw data was saved to the storage disk. The stability angle refers to the angle between the vertical projection line of the center of gravity and the line connecting the center of gravity to the ankle joint of the supporting foot, including the front stability angle and the rear stability angle. The trunk angle is the angle between the line connecting the midpoint of the left and right shoulder joints to the midpoint of the left and right hip joints and the sagittal axis. The hip angle refers to the angle between the trunk and the thigh. The knee angle refers to the angle between the thigh and the calf. The ankle angle refers to the angle between the calf and the sole of the foot. The push-off angle refers to the angle between the vector direction of the body's center of gravity to the rotation center point of the push-off leg's ankle joint and the sagittal axis. The glide angle refers to the angle between the line connecting the body's center of gravity to the rotation center point of the floating leg's ankle joint and the sagittal axis.

1.5 Data Analysis and Statistics

To better explain the changes in body posture during the athlete's starting process, the starting process is divided into three stages: the ready position (RD), the start (ST), and the sprinting (SP). The phase after the start is further divided into the touch-down moment (TD) and the take-off moment (TO) [4, 6, 7], as shown in Figure 2. The time taken to reach 15 meters is a key indicator for measuring the effectiveness of the start, with a shorter time indicating better effectiveness [3, 8]. Descriptive statistics (mean ± standard deviation) and Pearson correlation analysis were performed using SPSS 26.0 software for data statistics. Pearson correlation determines the relationship between variables (with P ≤ 0.05 set as the threshold), and the degree of correlation is defined (r): low correlation (0.1 to 0.3), moderate correlation (0.3 to 0.5), high correlation (0.5 to 0.7), very high correlation (0.7 to 0.9), and near-perfect (0.9 to 1.0).

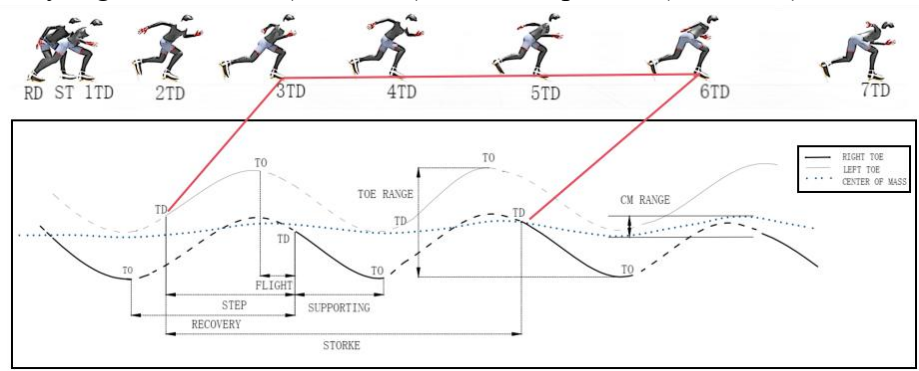

**Fingure 2 The Start Time or Phase of a Speed Skater**

## 2 Results

The technical characteristics of the athletes' starting stance and their correlation with start speed are shown in Table 1. The results indicate that the athletes' starting stance exhibits technical features such as a significantly larger rear stability angle compared to the front stability angle, with the hip, knee, and ankle angles of the rear leg being greater than those of the front leg. Additionally, the trunk angle and the angle of the push-off against the ice are smaller, the stride length is greater than the stride width, and the center of gravity is lower.

The rear stability angle (r=0.355), front leg knee angle (r=0.618), rear leg knee angle (r=0.459), and stride length (r=0.515) show a moderate or high positive correlation with start speed, while the trunk angle (r=-0.585) exhibits a high negative correlation with start speed. Other indicators do not correlate with start speed. The stepwise regression equation with start speed (y) as the dependent variable and the rear stability angle (a), trunk angle (b), left leg knee angle (c), right leg knee angle (d), and starting stance stride length (e) as independent variables is:

$$y=-1.58+0.026a-0.011b+0.034c-0.006d+1.072e$$

$(R^2=0.677, P<0.001)$

From the unstandardized coefficients, it is evident that the starting stance stride length, left leg knee angle, and rear stability angle have a greater impact on start speed.

**Table 1 Technical Characteristics of the Ready Position and its Correlation with the Starting Speed**

| Indicators | $\bar{x} \pm s$ | r |
|---|---|---|
| Stable angle（°） | 31.64±2.47 | 0.292 |
| Previous stable angle（°） | 6.98±3.56 | -0.272 |
| Post-stabilization angle（°） | 29.5±3.66 | 0.355* |
| Left hip angle（°） | 52.37±6.65 | 0.065 |
| Right hip angle（°） | 85.46±7.9 | 0.119 |
| Left knee angle（°） | 106.31±6.42 | 0.618** |
| Right knee angle（°） | 114.12±8.32 | 0.459** |
| Left ankle angle（°） | 89.18±3.92 | -0.073 |
| Right ankle angle（°） | 100.86±7.04 | 0.303 |
| Trunk angle（°） | 19.83±5.92 | -0.585** |
| Ice Sculling Angle(°) | 41.46±3.14 | -0.244 |
| Step width(m) | 0.31±0.48 | 0.107 |
| Step size(m) | 0.45±0.76 | 0.515** |
| Center of gravity height / height | 0.44±0.16 | 0.271 |

Note: * indicates "p<0.05", ** indicates "p<0.01"

The technical characteristics and indicators of the start and sprint technique, as well as their correlation with start speed (Table 2), show that the trunk angle gradually increases in the first 2-3 steps and then gradually decreases; the hip angle exhibits a technical feature where the right leg is significantly larger than the left leg from TD1 to TD7, and a trend of decreasing on the same side from TO1 to TO7, with the right hip angle still being larger than the left; the floating leg knee angle shows a trend of increasing on the same side in the first 6 steps of the sprint, with the right knee angle significantly larger than the left, and the right leg knee angle not reaching full extension at the TO1 moment, while the push-off leg knee angle shows a trend of increasing on the same side in subsequent TO moments, with a slight predominance on the right side; the floating leg skate angle at TD2 shows a significant increase in the right ankle angle, with subsequent TD or TO moments displaying a relatively stable change trend in the ankle angle, and a technical feature of the right ankle angle being higher than the left; the change in the floating leg ice contact angle after the start of the athlete is maintained between 77° and 80°, with a small variation, and the left leg ice contact angle is slightly higher than the right; the change in the push-off angle is maintained between 40° and 45°, with a significant decreasing trend starting from the TO7 moment.

The trunk angle shows a moderate negative correlation with start effectiveness at TO4 (r=-0.418), TD4 (r=-0.39), TO6 (r=-0.428), TD6 (r=-0.478) moments; the hip angle shows a moderate or high negative correlation with start effectiveness at TO1 (r=-0.352), TO4 (r=-0.447), TO6 (r=-0.531) moments; the knee angle shows a moderate positive correlation with start effectiveness at TD1 (r=0.488) moment; the ice contact angle shows a moderate positive correlation with start effectiveness at TD2 (r=0.4), TD4 (r=0.329), TD5 (r=0.385), TD6 (r=0.351) moments; the push-off angle shows a moderate positive correlation with start effectiveness at TO1 (r=0.419),

TO4 (r=0.421), TO7 (r=0.497) moments. Other indicators are not correlated with start effectiveness.

**Table 2 Technical characteristics of starting and sprinting and its correlation with starting effectiveness**

| Moment | Trunk angle (°) $\bar{x}\pm s$ | r | Hip angle (°) $\bar{x}\pm s$ | r | Knee angle (°) $\bar{x}\pm s$ | r | Ankle angle (°) $\bar{x}\pm s$ | r | Ice angle (°) $\bar{x}\pm s$ | r | Ice-skating corner (°) $\bar{x}\pm s$ | r |
|---|---|---|---|---|---|---|---|---|---|---|---|---|
| R-TO1 | 39.16±6.36 | -0.076 | 145.57±9.44 | -0.352* | 150.35±6.54 | -0.156 | 127.19±9.45 | -0.124 |  |  | 41.46±3.14 | 0.419* |
| L-TD1 |  |  | 75.99±9.04 | 0.256 | 94.88±6.1 | 0.488** | 90.79±4.64 | -0.183 | 76.77±3.81 | 0.004 |  |  |
| L-TO2 | 40.81±5.99 | -0.194 | 136.77±7.73 | -0.092 | 144.06±5.01 | 0.318 | 106.30±5.48 | -0.197 |  |  | 45.10±1.89 | 0.215 |
| R-TD2 | 42.52±6.06 | -0.139 | 86.89±9.43 | -0.043 | 108.9±5.62 | -0.001 | 109.06±5.43 | 0.228 | 77.34±6.01 | 0.400* |  |  |
| R-TO3 | 44.61±6.82 | -0.171 | 142.24±8.34 | -0.109 | 146.21±8.13 | 0.122 | 121.09±5.76 | 0.123 |  |  | 44.32±1.59 | 0.137 |
| L-TD3 | 46.4±6.51 | -0.219 | 79.69±10.05 | -0.121 | 104.58±5.8 | 0.217 | 99.86±6.03 | -0.123 | 79.98±2.82 | -0.163 |  |  |
| L-TO4 | 44.91±6.66 | -0.418** | 135.49±10 | -0.447** | 146.40±7.95 | 0.117 | 112.96±10.1 | 0.135 |  |  | 45.53±2.10 | 0.421* |
| R-TD4 | 46.45±6 | -0.39* | 89.54±9.82 | -0.194 | 111.2±5.55 | 0.053 | 107.38±7.60 | -0.306 | 77.24±4.58 | 0.329* |  |  |
| R-TO5 | 44.35±6.24 | -0.233 | 140.1±9.53 | -0.133 | 147.02±8.02 | 0.075 | 118.46±6.7 | 0.134 |  |  | 42.94±1.08 | 0.243 |
| L-TD5 | 45.03±6.02 | -0.205 | 78.43±8.53 | -0.144 | 107.00±5.18 | -0.019 | 104.45±5.41 | 0.184 | 78.52±2.6 | -0.385* |  |  |
| L-TO6 | 42.97±4.68 | -0.428** | 133.50±7.77 | -0.531** | 149.97±8.12 | -0.057 | 116.21±8.24 | -0.029 |  |  | 44.00±2.28 | -0.296 |
| R-TD6 | 43.57±5.2 | -0.478** | 86.20±8.63 | 0.008 | 115.17±5.43 | 0.201 | 106.68±5.79 | 0.018 | 76.72±5.16 | 0.351* |  |  |
| R-TO7 | 41.38±5.79 | -0.313 | 137.64±10.1 | -0.227 | 150.22±10.06 | 0.204 | 118.67±7.26 | -0.159 |  |  | 40.14±2.02 | 0.497** |
| L-TD7 | 40.42±5.48 | -0.312 | 74.07±7.30 |  | 106.9±3.73 | 0.159 | 103.28±5.62 | -0.074 | 79.35±4.09 | 0.191 |  |  |

Note: * indicates "p<0.05", ** indicates "p<0.01", R represents the right leg, and L represents the left leg.

## 3 Analysis and Discussion

### 3.1 The Influence of Body Posture Stability on Starting Performance

The key factors influencing the starting speed of short-track speed skaters include the stride length of the preparatory posture, the knee angle of the left leg, and the posterior stability angle. The results of this study indicate that, within a certain range, appropriately increasing the stride length, the knee angle of the left leg, and the posterior stability angle can enhance the starting speed. The stability of the starting posture is a prerequisite for ensuring an effective start and is also a critical factor determining the starting efficiency and sprint acceleration [9]. Due to the extremely low friction coefficient of the ice surface, speed skaters must start on ice while wearing skates with minimal contact area between the blade and the ice, resulting in poor stability of the support point and making it challenging to maintain a stable starting posture. Data analysis reveals that the anterior stability angle (6.89±3.56°) of athletes is significantly smaller than the posterior stability angle (29.50±3.66°), indicating that the starting posture places higher demands on posterior stability. Appropriately increasing the stride length and the knee angle of the left leg can effectively increase the posterior stability angle, thereby providing favorable conditions for the full extension of the rear push-off leg.

The ability to control single-leg support stability during the sprint acceleration phase is a crucial factor affecting starting performance. During the sprint, there is a contradiction between the forward shift of the athlete's center of gravity and the stability of single-leg sliding support. The sprint acceleration phase requires overcoming gravity and concentrating force on the single-leg skate, i.e., the single-leg ice contact posture. Additionally, the height between the skate blade and the sole (approximately 5.5–6.7 cm) increases the ankle joint's height from the ground, thereby extending the resistance arm length during the push-off phase and the moment of forward swing after push-off. Therefore, enhancing the athlete's ability to control stability during single-leg support is of paramount importance.

**3.2 The Effect of Forward Body Lean on Starting Performance**

The forward-leaning posture is the foundation for maintaining dynamic support and sliding balance, preventing athletes from leaning backward due to the torque generated by forward force components. From an aerodynamic perspective, a smaller trunk angle helps reduce the frontal area, thereby decreasing air resistance. During the starting acceleration phase, an appropriate trunk angle is more conducive to energy output, with the benefits far outweighing the energy loss caused by air resistance. Moreover, the counter-torque generated by the body's center of gravity during sliding provides balance support for the athlete. This unique push-off and sliding pattern results in the athlete's body exhibiting continuous sinusoidal motion on the ice. During the sprint acceleration phase, athletes need to reduce the ice contact angle to shift the center of gravity forward, thereby increasing the moment in the sliding direction. The greater the forward lean, the more force is transmitted backward, analogous to the principle of striking an object with a hammer from a height. This phenomenon provides insights for land-based specialized training, emphasizing the need to strengthen lower limb muscle extension training in a forward-leaning posture. During the sprint, the magnitude and direction of the force exerted by the skate on the ice vary with changes in trunk angle, the trajectory of the center of gravity, sliding direction, and speed. Unlike the fixed support characteristics of sprinters, speed skating starting techniques feature "dynamic support" or "sliding support." At the initial stage of the sprint, when the free-leg skate contacts the ice, the athlete's center of gravity is positioned in front of the contact foot, allowing smooth force application to the ice and minimizing braking force. However, if the center of gravity does not move past the skate's contact point in time, a "braking" effect will occur. Therefore, it is essential to establish a technical characteristic where the projection point of the center of gravity on the ice is located in front of the contact foot during the initial steps of the start.

**3.3 The Influence of Lower Limb Joint Extension Ability on Starting Performance**

Changes in lower limb joint angles significantly reflect the force characteristics and joint stiffness of the push-off leg [10, 11]. To generate force more effectively, athletes need to fully extend the push-off leg as much as possible. In short-distance events, the hip joint flexion-extension pattern plays a central role in the kinetic chain transmission during the starting phase. A larger extension range in a shorter time contributes to increased acceleration, representing the optimal external manifestation of starting performance. During the sprint acceleration phase after the start, a larger hip angle at the take-off moment is an indicator of good technical performance. During the start, the lower limb joint movement pattern of the rear support leg during the push-off phase is primarily related to extension. The kinetic chain transmission process rapidly propels the center of gravity from the proximal to the distal end through the extension of the rear support leg joints, reaching maximum speed at the moment of leaving the ice. The rear support leg achieves forward acceleration through rapid extension, and changes in lower limb joint angles directly reflect the effectiveness of the extension.

At the take-off moment during the start, the knee angle of the rear support leg is (150.34±6.54)°, and the inability to fully extend may be related to the athlete's eagerness to increase movement frequency. Although rapid knee extension during the support phase can generate greater power [12, 13], the effectiveness of the rear support leg's push-off is more critical than high power output during the initial starting phase [4]. Additionally, the unique hinge structure of Clap skates makes it difficult for athletes to achieve full extension of the rear support leg during the start. Coupled with the extremely low friction coefficient of the ice and the absence of a starting block, athletes can only rely on the inner edge of the rear support leg skate to embed into the ice and resist body weight for force generation. The small support area and poor stability make excessive explosive force likely to cause side slipping or "slip-out" phenomena. Therefore, the effectiveness of the push-off leg's extension is more critical for improving starting efficiency.

From the perspective of the human kinetic chain, the force for lower limb extension originates from the hip, with the kinetic chain transmission sequence being hip, knee, and ankle. If the hip and knee generate significant force while the ankle remains lax, it is difficult to obtain sufficient ice support reaction force. Furthermore, eversion of the push-off leg's ankle joint reduces starting efficiency. Therefore, strength training should focus on enhancing eversion control ability and rapid extension training with the ankle as the support, while integrating muscle contraction speed and movement amplitude to avoid functional strength imbalances and achieve the desired training effects.

## 3.4 The Influence of Ice Contact Support Ability and Push-Off Quality on Starting Performance

The rationality of the ice contact angle directly affects center of gravity transfer, push-off direction, and body stability, making it an important indicator of ice contact action quality [14]. As the race progresses and speed increases, the displacement of the athlete's body from the central axis gradually increases, and the sliding distance after the free-leg ice contact also extends, placing higher demands on the stability control ability of the skate's ice contact support. To maintain the sliding direction and ensure dynamic stability in a forward-leaning posture, the ankle plantar flexors and their associated muscle groups must possess strong extension control ability and firmly apply force to the inner edge of the skate. When the push-off leg is about to leave the ice, the Clap skate's heel automatically detaches from the boot, allowing full plantar flexion of the ankle and completing the push-off action. This places extremely high demands on the ankle's ability to control the skate. It is recommended that athletes perform land-based training in unstable conditions using sports shoes or skates, employing progressive training methods to enhance ankle strength and unstable support control ability, thereby improving "ice feel." Additionally, during specialized starting strength training, single-leg support end-release training and dynamic center of gravity transfer training should be emphasized to improve the effective connection between starting phases and the efficiency of force transmission between limbs.

The size of the push-off angle significantly affects the trajectory of the athlete's center of gravity and the effectiveness of push-off force. From an anatomical perspective, an excessively large push-off angle increases the vertical force component, raising the height of the center of gravity at the take-off moment and negatively affecting forward momentum, which is detrimental to acceleration. Conversely, an excessively small push-off angle causes over-plantar flexion of the ankle, hindering the exertion of push-off force [15]. Although a smaller push-off angle increases the forward force component, the reduction in vertical force may lead to premature ice contact of the free leg, also affecting speed. Multiple studies have shown that the push-off angle is a key indicator determining push-off effectiveness [4,10,16]. Minimizing the push-off angle during the early phase of the push-off leg's extension is beneficial for the vectorial exertion of horizontal force [17]. Applying "effective" force within the optimal push-off angle range is crucial for improving starting speed. During the sprint acceleration phase, the push-off force of the free-leg skate's ice contact support is characterized by instant acceleration and explosive knee extension. Therefore, training should focus on enhancing lower limb extension explosive power, particularly single-leg explosive power.

## 4 Conclusion

The starting preparatory posture places higher demands on the athlete's posterior stability. Appropriately increasing the posterior stability angle, knee angles of both legs, and stride length can enhance starting speed. During the starting and sprint phases, appropriately reducing the ice contact angle and push-off angle for each step while increasing the trunk angle and the range of hip angle changes at the take-off moment can improve starting efficiency. When optimizing starting posture training, it is necessary to appropriately reduce the trunk angle, increase stride length, forward shift of the center of gravity, and knee angle to better control starting stability and provide conditions for the rear push-off leg's force exertion. The training focus during the starting and sprint acceleration phases should be on strengthening lower limb joint angle control, minimizing the ice contact angle to avoid buffering effects, reducing the push-off angle to enhance effective push-off force, and reinforcing the ankle's "rigid support" function and the explosive power of the left leg.

## References：

Funding Project: Heilongjiang Provincial Natural Science Foundation Project in 2022 (Number: LH2022A018); Harbin Institute of Physical Education Doctoral Student Innovation Project in 2021 (Number: BS20-202107).